\documentclass[paper,12pt,nofootinbib,notitlepage]{revtex4-1}
\usepackage{graphicx}
\usepackage{amsmath}
\usepackage{amssymb}
\usepackage{dcolumn}
\usepackage{bm}
\usepackage{color}
\usepackage{dsfont}
\usepackage{feynmp}
\usepackage{marvosym}
\usepackage{phaistos}

\newcommand \beq{\begin{eqnarray}}
\newcommand \eeq{\end{eqnarray}}

\begin{document}
\unitlength=1mm
\allowdisplaybreaks

\title{SU($N$) Cartan-Weyl bases and color factors}

\author{Duifje Maria van Egmond}
\affiliation{Centre de Physique Th\'eorique, CNRS, Ecole polytechnique, IP Paris, F-91128 Palaiseau, France.}

\author{Urko Reinosa}
\affiliation{Centre de Physique Th\'eorique, CNRS, Ecole polytechnique, IP Paris, F-91128 Palaiseau, France.}

\date{\today}

\begin{abstract}
Constant temporal backgrounds have become a convenient tool for the discussion of continuum non-abelian gauge theories at finite temperature, as they provide a grasp on the confinement/deconfinement transition. It has been pointed that such backgrounds are better dealt with within so-called Cartan-Weyl color bases, rather than within the standard Cartesian bases. Here, we discuss the specificities of the evaluation of SU($N$) color factors within these bases, extending the discussion in Ref.~\cite{Reinosa:2015gxn}.
\end{abstract}

\maketitle

\section{Introduction}

It has become apparent that constant temporal gluonic backgrounds \cite{Fukushima:2003fw,Braun:2007bx,Dumitru:2012fw} offer an entry point into the physics of the confinement/deconfinement transition within continuum non-abelian gauge theories \cite{Fukushima:2003fw,Braun:2007bx,Braun:2009gm,Braun:2010cy,Dumitru:2012fw,Reinhardt:2012qe,Reinhardt:2013iia,Fischer:2013eca,Fischer:2014vxa,Reinosa:2014ooa,Herbst:2015ona,Reinosa:2015oua,Reinosa:2015gxn,Quandt:2016ykm,Maelger:2017amh,Reinhardt:2017pyr,Guo:2018scp,Maelger:2019cbk,Fu:2019hdw}. The constant temporal background $\bar A_\mu(x)=\bar A\,\delta_{\mu0}$ is determined from the minimization of the corresponding background effective potential\footnote{For a recent alternative proposal based on the standard effective potential, see \cite{VanEgmond:2021mlj}}.and offers a convenient alternative to the Polyakov loop as a way to monitor the transition as a function of the temperature.  Moreover, since the background couples to any colored field $|\varphi\rangle$ in a given representation $R$ through the covariant derivative
\beq
D_\mu|\varphi\rangle=\partial_\mu|\varphi\rangle-ig\bar A^at_R^a\delta_{\mu0}\,|\varphi\rangle\,,\label{eq:bc}
\eeq
the transition retroacts a priori on any other observable involving such fields.

One practical drawback of the coupling to the background in (\ref{eq:bc}) is that the different color modes in the representation $R$ are coupled to each other. It has been emphasized that the use of appropriate bases in the representation space simplifies the problem. First, without loss of generality, $\bar A$ can be color-rotated to lie in the commuting part of the algebra, the so-called Cartan subalgebra, $\bar A= \bar A^j t^j$ with $[t^j,t^{j'}]=0$. Second, one can decompose the state $|\varphi\rangle$ along a basis of states $|\rho\rangle$ that diagonalize simultaneously the action of the $t^j_R$, $t^j_R|\rho\rangle=\rho^j|\rho\rangle$.  The peculiarity of these bases  is that they label the various color modes in terms of vectors $\rho$, known as weights, whose components $\rho^j$ are the color charges associated to the commuting generators of the algebra. In such bases, Eq.~(\ref{eq:bc}) becomes
\beq
D_\mu\varphi^\rho=\partial_\mu\varphi^\rho-ig\bar A^j\rho^j\delta_{\mu0}\,\varphi^\rho\,,
\eeq
with $\varphi^\rho\equiv \langle\rho|\varphi\rangle$, meaning that the action of the covariant derivative is diagonal in color space. Switching to Fourier space, with the convention $\partial_\mu\to -iP_\mu$, the covariant derivative boils then down to the mere multiplication of $\varphi^\rho$ by a shifted momentum
\beq
P_\mu^\rho\equiv P_\mu+T(r\cdot \rho)\delta_{\mu 0}\,,\label{eq:shift}
\eeq
that combines the usual momentum $P_\mu$ and the charge $\rho$ carried by the field $\varphi^\rho$, where $r\cdot\rho\equiv r^j\rho^j$ and $r^j\equiv \beta g\bar A^j$.  The background $r^j$ appears then as a collection of (imaginary) chemical potentials associated to the various commuting charges $t^j$, thus facilitating both the interpretation and the practical implementation of the background field formalism. 

In the particular case of the adjoint representation, the representation space is the algebra itself, and one is lead to decompose any adjoint field $\varphi$ not along the usual Cartesian bases, $\varphi=\varphi^at^a$, but rather along so-called Cartan-Weyl bases, $\varphi=\varphi^\kappa t^\kappa$, whose elements satisfy $[t^j,t^\kappa]=\kappa^jt^\kappa$, with the adjoint weights $\kappa$  giving the charges in the adjoint representation associated to the commuting generators. The adjoint covariant derivative amounts again to the multiplication of $\varphi^\kappa$ by a shifted momentum of the form (\ref{eq:shift}) with $\rho$ replaced by $\kappa$. A specificity of the adjoint representation is that some of the charges $\kappa$ can be zero, in which case the shifted momentum coincides with the standard momentum. Details on the construction of Cartan-Weyl bases are provided in Sec.~\ref{sec:color}.

The benefits of the bases $\{|\rho\rangle\}$ and $\{t^\kappa\}$ are numerous. First, they greatly facilitate the identification of Weyl chambers and Weyl rotations (in the space spanned by the $r^j$), which are key elements in discussing the various symmetries of the problem in the background field formulation,\footnote{In particular, this allows to describe the center symmetry that relates to the confinement/deconfinement phase transition.} see for instance \cite{Reinosa:2020mnx} for a thorough discussion. They also make certain conservation rules explicit. In particular, the presence of a preferred color direction as defined by the background does not break the invariance under color rotations whose axis lies within the Cartan subalgebra. From Noether's theorem, this implies that the corresponding charges, which are nothing but the weights of the various representations, should be conserved. This conservation rule is visible at the level of the Feynman rules provided one uses the above mentioned bases. For instance, three shifted momenta $Q^\kappa$, $K^\lambda$ and $L^\tau$ associated to adjoint fields that meet at a vertex couple through a structure constant $f^{\kappa\lambda\tau}$ which is $0$ whenever the charges are not conserved at the vertex, that is $\kappa+\lambda+\tau\neq 0$ \cite{Reinosa:2015gxn}.\footnote{Similarly, three shifted momenta $Q^\kappa$, $K^\rho$ and $L^\sigma$ associated respectively to a gluon and to two fields in a representation $R$, couple through a factor $(t^\kappa_R)_{\rho\sigma}$ which vanishes whenever the charges are not conserved, that is $\rho\neq\sigma+\kappa$.} In turn, the conservation of the color charges at the vertices, once combined with the conservation of momenta, implies the conservation of the shifted momenta. This property allows one to export to the present context, some of the standard techniques  of the zero background formalism that are used to manipulate the Feynman integrals. In fact, owing to this feature, one can very often guess the results in the presence of a background from the corresponding results in the absence of background \cite{Reinosa:2015gxn}.

Despite all these simplifying features, one intricacy that arises due to (\ref{eq:shift}) is that the color factors are inevitably mixed with the Feynman integrals and it is in general not possible to evaluate the corresponding color traces independently of the evaluation of the integrals. There are some situations, however, where it is still possible to do so to some extent. One example arises from the thermal decomposition of the Feynman integrals \cite{Blaizot:2004bg,vanEgmond:2020lui}. In this type of decomposition, one writes a given loop integral at finite temperature as a sum of contributions where certain loops correspond to zero-temperature amplitudes. For all the momenta within these amplitudes, the shift in (\ref{eq:shift}) vanishes and thus, the corresponding color charges appear only within the color factors. Also, in some models \cite{Kroff:2018ncl}, it can happen that some of the color charges appear only in the structure constants and not as shifts of the momenta, even in the case of purely thermal contributions. In these two situations, it is possible to sum over the corresponding color labels within the color factors. The calculations are pretty similar to those in the standard Cartesian bases provided some care is taken to account for the specificities of the Cartan-Weyl bases. We discuss these questions in Secs.~\ref{sec:struc} and \ref{sec:traces} including the treatment of the structure constants $f^{\kappa\lambda\tau}$ and $d^{\kappa\lambda\tau}$.

In the generic situation where the color labels are also present within the shifted momenta, there is nothing one can do a priori, except when some of the color labels that are summed over correspond to zero charges, such as in the case of the adjoint charges alluded to above. For such zero color labels, the shift in (\ref{eq:shift}) is again zero, and one can perform a partial summation over the zero labels within the color factor. We discuss this type of partial color summation in Sec.~\ref{sec:zeros} while providing more insight on the tensors $f^{\kappa\lambda\tau}$ and $d^{\kappa\lambda\tau}$.\\

We shall limit our analysis to the su($N$) Lie algebras for their obvious application to the description of the fundamental interactions. However, some of the discussion can be extended to any Lie algebra admitting Cartan-Weyl bases.

\section{The su($N$) algebra and its Cartan-Weyl bases}\label{sec:color}
We consider SU($N$), the group  of special unitary matrices of size $N$. This is a Lie group whose Lie algebra, denoted su($N$), is the space of traceless anti-hermitian matrices of size $N$. Here, we follow the mathematicians convention that relates the group elements $U\in SU(N)$ and the algebra elements $X\in \mbox{su($N$)}$ as $U=e^X$. However, in order to keep contact with the physicists convention, we shall write any basis of the algebra as $\{it^a\}$ such that $U=e^{iX^at^a}$. We mention that, as any Lie algebra associated to a Lie group, su($N$) is a real algebra, meaning that any element $X\in\mbox{su($N$)}$ writes as a linear combination of the basis elements $it^a$ with real coefficients $X^a$. As we recall below, it is convenient to complexify this real algebra by allowing for complex coefficients $X^a$. The complexified su($N$) algebra is the space of all complex traceless matrices of size $N$.

\subsection{Standard basis}
A possible basis for the su($N$) algebra is $\{it^a\}=\{iH^j,iX^{jj'},iY^{jj'}\}$ where the (hermitian) matrices $H^j$, $X^{jj'}$ and $Y^{jj'}$ are defined in terms of their components
\beq
H^j_{kk'} & = & \rho^j_k\,\delta_{kk'}\,,
\eeq
with
\beq\label{eq:rho}
\rho^j_k=\frac{1}{\sqrt{2j(j+1)}}\times\left\{
\begin{array}{rl}
\!1\,,\,\, & \mbox{if $k\leq j$}\\
\!-j\,,\,\, & \mbox{if $k=j+1$}\\
\!0\,,\,\, & \mbox{if $k>j+1$}
\end{array}
\right.,
\eeq
for $1\leq j\leq N-1$, and
\beq
X^{jj'}_{kk'}=\frac{1}{2}\,\big(\delta_{kj}\delta_{j'k'}+\delta_{kj'}\delta_{jk'}\big) \quad \mbox{and} \quad Y^{jj'}_{kk'}=-\frac{i}{2}\,\big(\delta_{kj}\delta_{j'k'}-\delta_{kj'}\delta_{jk'}\big)\,,\label{eq:X}
\eeq
for $1\leq j<j'\leq N$. The pre-factors in Eqs.~(\ref{eq:rho})-(\ref{eq:X}) have been chosen such that ${\rm tr}\,t^at^b=\delta^{ab}/2$. There are $N-1$ matrices $H^j$, $N(N-1)/2$ matrices $X^{jj'}$ and $N(N-1)/2$ matrices $Y^{jj'}$. This gives a total of $N^2-1$ matrices, in agreement with the dimension of the su($N$) algebra. For $N=2$ and $N=3$, one recovers the Pauli and Gell-Mann matrices respectively (to within a factor $1/2$).

\subsection{Representations and weights}
The matrices $H^j$ are diagonal. For a given $k$, the diagonal elements $H^j_{kk}=\rho_k^j$ (with $1\leq j\leq N-1$) are eigenvalues that correspond to the same eigenstate and it is convenient to combine them into a vector $\rho_k$ of $\mathds{R}^{N-1}$ known as weight. More precisely, this defines a weight of the defining representation of su($N$),\footnote{We refrain from using the term ``fundamental representation'' as there are in fact $N-1$ fundamental representations for su($N$), of which the defining representation considered here is just one example.} or defining weight of su($N$). The collection of defining weights $\rho_k$ (with $1\leq k\leq N$) forms a structure in $\mathds{R}^{N-1}$ known as the weight diagram of the defining representation. 

\begin{figure}[t]
\begin{center}
\includegraphics[height=0.42\textheight]{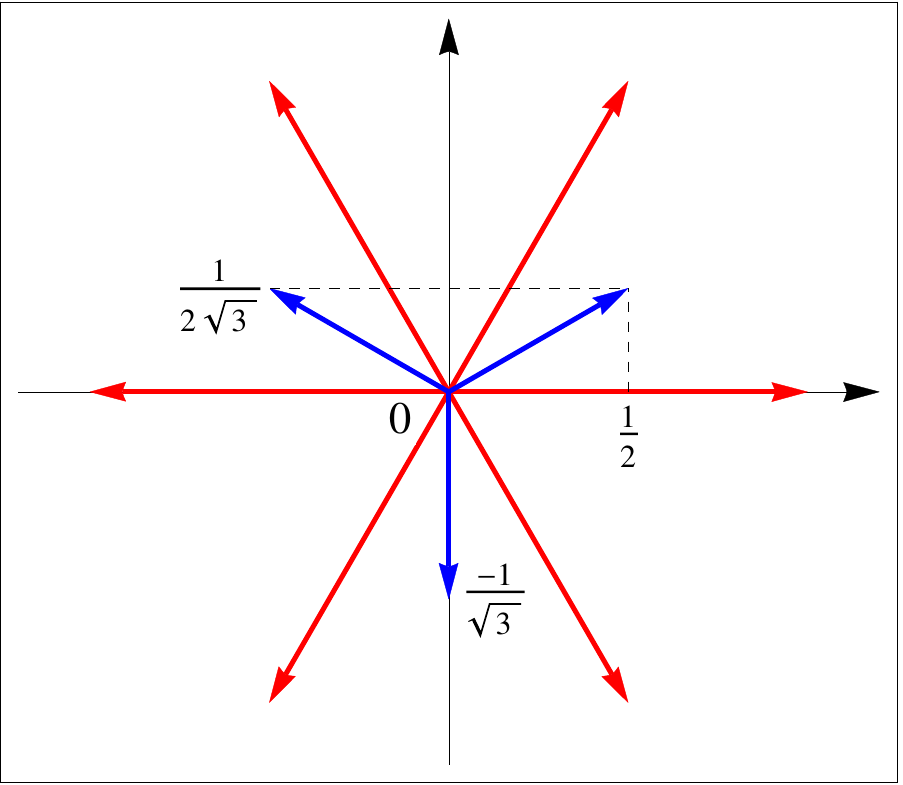}
\caption{Blue: weight diagram for the defining representation of su($3$). Red: root diagram of su($3$) (formed by the non-vanishing adjoint weights).}
\label{fig:roots}
\end{center}
\end{figure}

In the case of su($2$), the defining weights are vectors of $\mathds{R}$. They are obtained from $\sigma_3/2$ and correspond to $\pm 1/2$. Similarly, the defining weights of su($3$) are vectors of $\mathds{R}^2$. They are obtained from $\lambda_3/2$ and $\lambda_8/2$ and are represented in Fig.~\ref{fig:roots}. In general, the defining weights of su($N$) obey the following geometrical constraints
\beq
\rho_k^2\equiv\rho_k\cdot\rho_k=\frac{1}{2}\left(1-\frac{1}{N}\right)\, \quad \mbox{and} \quad \rho_k\cdot\rho_{k'}=-\frac{1}{2N} \quad \mbox{for any $k$ and $k\neq k'$}\,,\label{eq:constraints}
\eeq
which one easily deduces from Eq.~(\ref{eq:rho}). From these properties, one can show that any subset of defining weights, with the exception of the complete weight diagram, is a linearly independent set, see App.~\ref{app:sum} for more details. The complete weight diagram is not a linearly independent set because it is made of $N$ vectors in the $(N-1)$-dimensional space $\mathds{R}^{N-1}$. In fact, since the $H^j$ are traceless, one easily finds that the $N$ defining weights are constrained by
\beq
\sum_{k=1}^N \rho_k=0\,,
\eeq
which is in fact the only constraint among the defining weights.

Since the $H^j$'s form a complete set of commuting observables (CSCO), one can unambiguously label the eigenstates in terms of the defining weights as $|\rho\rangle$. By definition,
\beq
H^j\,|\rho\rangle=\rho^j\,|\rho\rangle\,,\label{eq:diag}
\eeq 
and, because the $H^j$ are hermitian, one can choose the $|\rho\rangle$'s such that
\beq
\langle\rho|\sigma\rangle=\delta^{\rho\sigma}\,,
\eeq
where the Kronecker delta $\delta^{\rho\sigma}$ is defined to be equal to $1$ if $\rho=\sigma$ and $0$ otherwise. The states $|\rho\rangle$ form a basis of the defining representation space that can be used to decompose the color structure of various tensors. In particular, any state $|X\rangle$ in the defining representation space writes
\beq
|X\rangle=\sum_\rho X^\rho\,|\rho\rangle\,,\label{eq:sum}
\eeq
with $X^\rho=\langle \rho|X\rangle$.\\

Most of the previous considerations extend to any finite representation $R$ of the su($N$) algebra over a complex vector space. Without loss of generality, one can assume that the $H^j_R$ are hermitian, and look for the eigenstates that diagonalize them simultaneously. The eigenvalues associated to a given eigenstate can again be combined into a vector $\rho$ called weight of the representation $R$. The collection of all such weights forms a structure in $\mathds{R}^{N-1}$ known as the weight diagram of the representation $R$. We note that the weights of a given representation $R$ do not need to obey the same geometrical constraints as those of the defining representation, see Eq.~(\ref{eq:constraints}).

 Another potential difference with respect to the defining representation is that the $H^j_R$ may not form a CSCO: a given weight $\rho$ could appear multiple times as the weight associated to various independent eigenstates. In this case, a more accurate notation for the eigenstates would be $|\rho^{(\mu)}\rangle$ where $(\mu)$ labels the multiplicity of the weight $\rho$ and is used to distinguish various eigenstates sharing this same weight. In other words, for $\mu\neq\nu$, we have $|\rho^{(\mu)}\rangle\neq |\rho^{(\nu)}\rangle$ but $\rho^{(\mu)}=\rho^{(\nu)}=\rho$. By definition one has\footnote{One should clearly distinguish $\rho^j$ from $\rho^{(\mu)}$. The former is just the $j$-th component of the latter.}
\beq
H^j_R\,|\rho^{(\mu)}\rangle=\rho^j\,|\rho^{(\mu)}\rangle\,,\label{eq:diag2}
\eeq
and, because the $H^j_R$'s are hermitian, one can choose the $|\rho^{(\mu)}\rangle$'s such that
\beq
\langle\rho^{(\mu)}|\sigma^{(\nu)}\rangle=\delta^{\rho\sigma}\delta^{\mu\nu}\,.
\eeq
The states $|\rho^{(\mu)}\rangle$ form a basis of the representation space that can be used to decompose the color structure of various tensors. In particular, any state $|X\rangle$ in the representation space writes
\beq
|X\rangle=\sum_{\rho\,\mu} X^{\rho^{(\mu)}}\,|\rho^{(\mu)}\rangle\,, 
\eeq
with $X^{\rho^{(\mu)}}=\langle\rho^{(\mu)}|X\rangle$.

In what follows, to avoid notation cluttering, we shall adopt the following simplifying rule: even in the case where some of the weights have a multiplicity different from $1$, we shall denote a generic weight of the representation as $\rho$ and only when considering a specific weight, say $\rho_0$, shall we indicate its multiplicity label as $\rho_0^{(\mu)}$. This means in practice that we can use expressions such as (\ref{eq:diag}) or (\ref{eq:sum}) in the general case (of course with $H^j$ replaced by $H^j_R$ and $|X\rangle$ taken in the appropriate representation space). One just needs to pay attention to the meaning of certain tensors. In particular, with this notational convention, $\delta^{\rho\sigma}$ is non-zero iff the weights associated to $\rho$ and $\sigma$ as well as their multiplicity labels coincide, as the more explicit (but also more cumbersome) notation $\delta^{\rho_0^{(\mu)}\sigma_0^{(\nu)}}=\delta^{\rho_0\sigma_0}\delta^{\mu\nu}$ would indicate. We also mention that, in this simplified notation, linear combinations such as $\rho_1+\rho_2$ refer to the sum of the corresponding weights, without any reference to the multiplicity labels. We shall see a concrete application of this convenient simplifying notation in the next section.

\subsection{Adjoint representation and Cartan-Weyl bases}
A particularly important representation is the adjoint representation $R={\rm ad}$. In this case, the space of the representation is the algebra itself and the simultaneous diagonalization of the $H^j_{\rm ad}$ amounts to constructing a basis of the algebra whose elements $X$ have commutators $[H^j,X]$ proportional to $X$. Such bases are known as Cartan-Weyl basis. In the case of su($N$), an example of Cartan-Weyl basis is provided by $\{H^j,Z^{jj'}\}$ where the matrices $H^j$ and $Z^{jj'}$ are defined by the components\footnote{We note that $Z^{jj'}=(X^{jj'}+iY^{jj'})/\sqrt{2}$ if $j<j'$ and $Z^{jj'}=(X^{jj'}-iY^{jj'})/\sqrt{2}$ if $j>j'$.}
\beq
H^j_{kk'} & = & \rho^j_k \delta_{kk'}\,, \quad 1\leq j\leq N-1\,,
\eeq
and
\beq
Z^{jj'}_{kk'} & = & \frac{1}{\sqrt{2}}\delta_{kj}\delta_{j'k'}\,, \quad  1\leq j,j'\leq N\,, \quad j\neq j'\,.
\eeq
That the action of the $H^j_{\rm ad}$ is diagonal on the elements of the basis $\{H^j,Z^{jj'}\}$ is clear from the following calculations (the first of which is trivial)
\beq
\big[H^j,H^{j'}\big]_{kk''} & = & \rho^j_k\delta_{kk'}\rho^{j'}_{k'}\delta_{k'k''}-\rho^{j'}_k\delta_{kk'}\rho^j_{k'}\delta_{k'k''}\nonumber\\ 
& = & (\rho^j_k\rho^{j'}_k-\rho^{j'}_k\rho^{j}_k)\delta_{kk''}=0\,,\label{eq:z}\\
\big[H^j,Z^{j'j''}\big]_{kk''} & = &  \frac{1}{\sqrt{2}}\Big(\rho^j_k\delta_{kk'}\delta_{k'j'}\delta_{j''k''}-\delta_{kj'}\delta_{j''k'}\rho^{j}_{k'}\delta_{k'k''}\Big)\nonumber\\
& = &  \frac{1}{\sqrt{2}}\Big(\rho^j_k\delta_{kj'}\delta_{j''k''}-\delta_{kj'}\delta_{j''k''}\rho^{j}_{k''}\Big)\nonumber\\
& = &  \frac{1}{\sqrt{2}}(\rho^j_{j'}-\rho^j_{j''})\delta_{kj'}\delta_{j''k''}=(\rho^j_{j'}-\rho^j_{j''})Z^{j'j''}_{kk''}\,,\label{eq:diff}
\eeq
which should be read as Eq.~(\ref{eq:diag2}) in the particular case of the adjoint representation, the $H^{j'}$ and $Z^{j'j''}$ playing the role of the eigenstates diagonalizing the action of the $H^j_{\rm ad}$'s, and the right-hand side of these equations granting access to the weights of the adjoint representation, or adjoint weights.

Let us mention that $(Z^{jj'})^\dagger =Z^{j'j}\neq Z^{jj'}$. It follows that $\{H^j,Z^{jj'}\}$ is not a basis of the original algebra (seen as a real algebra) but rather a basis of the complexified su($N$) algebra.  The need for considering the complexified algebra originates in that, despite commuting with each other, the $H^j_{\rm ad}$ have no reason to be diagonalizable within the original space of the representation, that is the real Lie algebra su($N$). On the other hand, it turns out that the $H^j_{\rm ad}$ are self-adjoint operators over the complexified su($N$) algebra, with respect to the inner product ${\rm tr}\,X^\dagger Y$,\footnote{Indeed ${\rm tr}\,[H^j,X]^\dagger\,Y={\rm tr}\,[X^\dagger,H^j]\,Y={\rm tr}\,[H^j,Y]\,X^\dagger={\rm tr}\,X^\dagger\,[H^j,Y]\,.$} thus ensuring their diagonalizability over this space. The diagonalization basis can be chosen orthonormal and, in fact, the basis $\{H^j,Z^{jj'}\}$ is orthonormal for the inner product $2\,{\rm tr}\,X^\dagger Y$, as a simple calculation reveals. There is of course no problem in working with bases of the complexified algebra, even though the basis elements are not in the original (real) su($N$) algebra. This is because any element of the original algebra can be obtained from a linear combination of these basis elements upon certain restrictions on the coefficients of the basis decomposition. Here, the only restriction is that while the coefficient of $H^j$ is real, the coefficients of $Z^{jj'}$ and $Z^{j'j}$ should be complex conjugate of each other.

Returning to Eqs.~(\ref{eq:z})-(\ref{eq:diff}), we find that there are two types of adjoint weights, vanishing and non-vanishing ones. The non-vanishing adjoint weights are usually called roots. They have multiplicity equal to $1$ since they are associated to a single eigenstate and, from Eq.~(\ref{eq:diff}), we see that they correspond to differences of non-equal defining weights, $\alpha_{j'j''}=\rho_{j'}-\rho_{j''}$, with $j'\neq j''$. There are thus as many roots as there are pairs of non-equal defining weights, that is $N(N-1)$. The roots also come by pairs since for any root $\alpha_{j'j''}$, $\alpha_{j''j'}=-\alpha_{j'j''}$ is also a root. The roots are generically written $\alpha$. Sometimes, we will need to access the two weights associated to a given root $\alpha$. We shall denote them as $\rho^\alpha$ and $\bar\rho^\alpha$ such that $\alpha=\rho^\alpha-\bar\rho^\alpha$. As for the vanishing adjoint weights, there is no conventional terminology or notation. We find it convenient to refer to them as zeros and to denote them $0^{(j')}$, the multiplicity label $j'$ referring to the fact that there are as many zeros as there are eigenstates $H^{j'}$, that is $N-1$.

It is conventional to label the elements of the Cartan-Weyl basis that have non-zero adjoint weights as $t^\alpha$ with $\alpha$ the corresponding roots. If we also set $t^j\equiv H^j$, the Cartan-Weyl basis rewrites $\{t^j,t^\alpha\}$ and the commutation relations (\ref{eq:z})-(\ref{eq:diff}) take the form $[t^j,t^{j'}]=0$ and $[t^j,t^\alpha]=\alpha^jt^\alpha$. We find it convenient to use a notation more in line with the general discussion in the previous section. We label any element of the Cartan-Weyl basis as $t^\kappa$ where $\kappa$ is the corresponding weight (including possible multiplicity labels), that is $\kappa=\alpha$ in the case where the basis element is associated to a root and $\kappa=0^{(j)}$ in the case where the basis element is associated to a (multiple) zero. With this notation, the Cartan-Weyl basis writes $\{t^\kappa\}$ and the commutation relations (\ref{eq:z})-(\ref{eq:diff}) condense into
\beq
[t^{0^{(j)}},t^\kappa]=\kappa^j t^\kappa\,,
\eeq
where $t^{0^{(j)}}$ is nothing but $H^j$. With these notations, one also easily verifies that 
\beq
(t^\kappa)^\dagger=t^{-\kappa} \quad \mbox{and} \quad {\rm tr}\,t^{-\kappa} t^\lambda=\frac{1}{2}\delta^{\kappa\lambda}\,,
\eeq 
where we stress again that the Kronecker delta should be interpreted with care since it is only non-zero (and equal to $1$) if the weights associated to $\kappa$ and $\lambda$, as well as their possible multiplicity labels, coincide. Those cases correspond either to $\kappa$ and $\lambda$ being the same root, $\kappa=\lambda=\alpha$, or $\kappa$ and $\lambda$ being a zero with the same multiplicity label, $\kappa=\lambda=0^{(j)}$.

\subsection{Fierz identity}
As we have already mentioned, $\{t^\kappa\}$ is a basis of the complexified su($N$) algebra. This means that linear combinations of the $t^\kappa$ with complex coefficients allow to generate any complex traceless matrix of size $N$. By adding the identity matrix $\mathds{1}$ to the basis $\{t^\kappa\}$, one can in fact generate any complex matrix $M$ of size $N$:
\beq\label{eq:C1}
M=m_0\,\mathds{1}+\sum_\kappa m_\kappa\,t^{-\kappa}\,.
\eeq
We have conventionally chosen to write $m_\kappa t^{-\kappa}$ rather than $m_\kappa t^\kappa$ for later convenience. Using that ${\rm tr}\,t^\kappa=0$ and ${\rm tr}\,t^{-\kappa} t^\lambda=\delta^{\kappa\lambda}/2$, one finds
\beq\label{eq:C2}
m_0=\frac{1}{N}\,{\rm tr}\,M\,, \quad m_\kappa=2\,{\rm tr}\,Mt^\kappa\,,
\eeq
and then
\beq\label{eq:C3}
M=\frac{1}{N}({\rm tr}\,M)\,\mathds{1}+\sum_\kappa(2\,{\rm tr}\,Mt^\kappa)\,t^{-\kappa}\,.
\eeq
The content of Eq.~(\ref{eq:C3}) can be equivalently summarized by choosing $M$ to be the matrix $Z^{kl}$ (including the possibility that $k=l$) such that $Z^{kl}_{ij}=\delta_{ik}\delta_{lj}$. In this case, Eq.~(\ref{eq:C3}) gives
\beq
\delta_{ik}\delta_{lj}=\frac{1}{N}\delta_{ij}\delta_{kl}+2\,\sum_\kappa t^{-\kappa}_{ij}\,t^\kappa_{lk}\,,
\eeq
which we rewrite as
\beq
\sum_\kappa t^{-\kappa}_{ij}\,t^\kappa_{lk}=\frac{1}{2}\delta_{ik}\delta_{lj}-\frac{1}{2N}\delta_{ij}\delta_{kl}\,.\label{eq:Fierz}
\eeq
This is nothing but the Fierz identity expressed in terms of the Cartan-Weyl generators $t^\kappa$. An immediate consequence is obtained by setting $\smash{j=l}$ and summing over $j$: one finds that $\sum_\kappa t^{-\kappa} t^\kappa$ is proportional to the identity, with a pre-factor, or {\it Casimir}, equal to $(N^2-1)/(2N)$. We shall exploit the Fierz identity further in the next sections.

\section{Structure constants $f^{\kappa\lambda\tau}$ and $d^{\kappa\lambda\tau}$}\label{sec:struc}

A particularly useful application of (\ref{eq:C3}) is obtained by choosing $M$ to be the matrix $t^\kappa t^\lambda$. One obtains
\beq\label{eq:C6}
t^\kappa t^\lambda=m^{\kappa\lambda}_{\,\,\,\,\,\,\,0}\,\mathds{1}+\sum_\tau m^{\kappa\lambda}_{\,\,\,\,\,\,\,\tau}\,t^{-\tau}\,.
\eeq
with
\beq
m^{\kappa\lambda}_{\,\,\,\,\,\,0}=\frac{1}{2N}\delta^{\kappa(-\lambda)}\,, \quad m^{\kappa\lambda}_{\,\,\,\,\,\,\tau}=2\,{\rm tr}\,t^\kappa t^\lambda t^\tau\,,
\eeq
where, owing to the cyclicality of the trace, $m^{\kappa\lambda}_{\,\,\,\,\,\,\tau}$ is invariant under cyclic permutations of $\kappa$, $\lambda$ and $\tau$. It is convenient to decompose this tensor into a symmetric and an anti-symmetric part with respect to the first two indices, which are then completely symmetric and completely anti-symmetric respectively. One writes $m^{\kappa\lambda}_{\,\,\,\,\,\,\tau}=(d^{\kappa\lambda\tau}+f^{\kappa\lambda\tau})/2$, with
\beq
d^{\kappa\lambda\tau}=2\,{\rm tr}\,\{t^\kappa, t^\lambda\} t^\tau\,, \quad f^{\kappa\lambda\tau}=2\,{\rm tr}\,[t^\kappa, t^\lambda]\,t^\tau\,.\label{eq:C9}
\eeq
Equation (\ref{eq:C6}) becomes then
\beq\label{eq:CC}
t^\kappa t^\lambda = \frac{1}{2N}\delta^{\kappa(-\lambda)}\mathds{1}+\frac{1}{2}\sum_\tau (d^{\kappa\lambda\tau}+f^{\kappa\lambda\tau})\,t^{-\tau}\,.
\eeq
The $f^{\kappa\lambda\tau}$'s are of course the structure constants in the basis $\{t^\kappa\}$
\beq
[t^\kappa,t^\lambda]=\sum_\tau f^{\kappa\lambda\tau}t^{-\tau}\,,\label{eq:DD}
\eeq 
(we note however the absence of the factor $i$ in our convention) while 
\beq
\{t^\kappa,t^\lambda\}=\frac{1}{N}\delta^{\kappa(-\lambda)}\mathds{1}+\sum_\tau d^{\kappa\lambda\tau}t^{-\tau}\,.\label{eq:EE}
\eeq
By taking the hermitian conjugate of Eq.~(\ref{eq:C9}) while using $(t^\kappa)^\dagger=t^{-\kappa}$, we find that
\beq\label{eq:cc}
(d^{\kappa\lambda\tau})^*=d^{(-\kappa)(-\lambda)(-\tau)}\,, \quad (f^{\kappa\lambda\tau})^*=-f^{(-\kappa)(-\lambda)(-\tau)}\,.
\eeq

\subsection{Jacobi identities}
The tensors $d^{\kappa\lambda\tau}$ and $f^{\kappa\lambda\tau}$ obey various identities, known as Jacobi identities. To derive them in full generality, let us start writing the following collection of products of three generators
\beq
t^\kappa t^\lambda t^\tau \quad  t^\lambda t^\kappa t^\tau \quad  t^\tau t^\kappa t^\lambda  \quad  t^\tau t^\lambda t^\kappa \nonumber\\
t^\lambda t^\tau t^\kappa \quad  t^\tau t^\lambda t^\kappa \quad  t^\kappa t^\lambda t^\tau  \quad  t^\kappa t^\tau t^\lambda\nonumber\\
t^\lambda t^\tau t^\kappa \quad  t^\tau t^\lambda t^\kappa \quad  t^\kappa t^\lambda t^\tau  \quad  t^\kappa t^\tau t^\lambda\nonumber
\eeq
where the last two lines are obtained by cyclically permuting the color indices $\kappa$, $\lambda$ and $\tau$ of the first line. By associating $+1$ or $-1$ to each of the terms we can find various ways of cancelling the terms. There are only four different possibilities that can be expressed in terms of commutators or anti-commutators. They write
\beq
\begin{array}{ccccccc}
[[t^\kappa,t^\lambda],t^\tau] & + & [[t^\lambda,t^\tau],t^\kappa] & + & [[t^\tau,t^\kappa],t^\lambda] & = & 0\,,\\
{[}\{t^\kappa,t^\lambda\},t^\tau{]} & + &[\{t^\lambda,t^\tau\},t^\kappa] & + & [\{t^\tau,t^\kappa\},t^\lambda] & = & 0\,,\\
{[}\{t^\kappa,t^\lambda\},t^\tau{]} & - & \{[t^\lambda,t^\tau],t^\kappa\} & + & \{[t^\tau,t^\kappa],t^\lambda\} & = & 0\,,\\
{[}[t^\kappa,t^\lambda],t^\tau{]} & - & \{\{t^\lambda,t^\tau\},t^\kappa\} & + & \{\{t^\tau,t^\kappa\},t^\lambda\} & = & 0\,.
\end{array}
\eeq
Using Eqs.~(\ref{eq:DD}) and (\ref{eq:EE}), these identities rewrite
\beq
f^{\kappa\lambda(-\eta)}f^{\tau\sigma\eta}+f^{\lambda\tau(-\eta)}f^{\kappa\sigma\eta}+f^{\tau\kappa(-\eta)}f^{\lambda\sigma\eta} & = & 0\,,\label{eq:GG}\\
d^{\kappa\lambda(-\eta)}f^{\tau\sigma\eta}+d^{\lambda\tau(-\eta)}f^{\kappa\sigma\eta}+d^{\tau\kappa(-\eta)}f^{\lambda\sigma\eta} & = & 0\,,\label{eq:LL}\\
d^{\kappa\lambda(-\eta)}f^{\tau\sigma\eta}-f^{\lambda\tau(-\eta)}d^{\kappa\sigma\eta}+f^{\tau\kappa(-\eta)}d^{\lambda\sigma\eta} & = & 0\,,\label{eq:II}
\eeq
as well as\footnote{Note that there is an overall sign with respect to a similar formula obtained using Cartesian color bases, see for instance \cite{su3}. This is because, our convention for the structure constants $f^{\kappa\lambda\tau}$ does not include a factor $i$, see Eq.~(\ref{eq:DD}).}
\beq\label{eq:KK}
f^{\kappa\lambda(-\eta)}f^{\tau\sigma\eta}=-\left[\frac{2}{N}\Big(\delta^{\kappa(-\tau)}\delta^{\lambda(-\sigma)}-\delta^{\kappa(-\sigma)}\delta^{\lambda(-\tau)}\Big)+d^{\kappa\tau(-\eta)}d^{\lambda\sigma\eta}-d^{\kappa\sigma(-\eta)}d^{\lambda\tau\eta}\right].
\eeq
In writing these equations, we have used a convention of summation over repeated color labels which we shall use systematically in what follows, unless specifically stated. The previous identities shall play a pivotal role in determining color traces in Sec.~\ref{sec:traces}.

\subsection{Special identity in the SU($3$) case}
In the case of SU($3$), there is an additional relation that one can derive between $f^{\kappa\lambda\tau}$ and $d^{\kappa\lambda\tau}$, which we obtain by extending the discussion of \cite{su3} to the Cartan-Weyl bases. Take $M$ to be a traceless hermitian matrix. In this case,  Eq.~(\ref{eq:C1}) writes
\beq
M=\sum_\kappa m_\kappa t^{-\kappa}\,,
\eeq
with $m_\kappa^*=m_{-\kappa}$. Such a matrix can be diagonalised and we denote its eigenvalues as $\mu_1$, $\mu_2$ and $\mu_3$. Moreover, the matrix $M$ nullifies its characteristic polynomial
\beq
M^3-(\mu_1+\mu_2+\mu_3)\,M^2+(\mu_1\mu_2+\mu_2\mu_3+\mu_3\mu_1)\,M-\mu_1\mu_2\mu_3=0\,.
\eeq
Since $M$ is traceless, we have $\mu_1+\mu_2+\mu_3={\rm tr}\, M=0$, and so
\beq
\mu_1\mu_2+\mu_2\mu_3+\mu_3\mu_1 & = & \frac{1}{2}\Big[(\mu_1+\mu_2+\mu_3)^2-\mu_1^2-\mu_2^2-\mu_3^2\Big]\nonumber\\
& = & -\frac{1}{2}(\mu_1^2+\mu_2^2+\mu_3^2)=-\frac{1}{2}{\rm tr}\,M^2\,,
\eeq
as well as
\beq
\mu_1\mu_2\mu_3 & = & \frac{1}{6}\Big[(\mu_1+\mu_2+\mu_3)^3-\mu_1^3-\mu_2^3-\mu_3^3\nonumber\\
& & \hspace{1.0cm} -\,3\mu_1^2(\mu_2+\mu_3)-3\mu_2^2(\mu_3+\mu_1)-3\mu_3^2(\mu_1+\mu_2)\Big]\nonumber\\
& = & \frac{1}{6}\Big[(\mu_1+\mu_2+\mu_3)^3+2(\mu_1^3+\mu_2^3+\mu_3^3)-3(\mu_1^2+\mu_2^2+\mu_3^2)(\mu_1+\mu_2+\mu_3)\Big]\nonumber\\
& = & \frac{1}{3}(\mu_1^3+\mu_2^3+\mu_3^3)=\frac{1}{3}{\rm tr}\,M^3\,.
\eeq
Now
\beq
{\rm tr}\,M^2=m_\kappa m_\lambda {\rm tr}\, t^{-\kappa}t^{-\lambda}=\frac{1}{2}m_\kappa m_{-\kappa}
\eeq
and
\beq
{\rm tr}\,M^3 & = & m_\kappa m_\lambda m_\tau\,{\rm tr}\, t^{-\kappa}t^{-\lambda}t^{-\tau}\nonumber\\
& = & \frac{1}{2}m_\kappa m_\lambda m_\tau\,{\rm tr}\, \{t^{-\kappa},t^{-\lambda}\}t^{-\tau}=\frac{1}{4}m_\kappa m_\lambda m_\tau\,d^{(-\kappa)(-\lambda)(-\tau)}\,.
\eeq
Thus
\beq
M^3-\frac{1}{4}m_\kappa m_{-\kappa}\,M-\frac{1}{12}m_\kappa m_\lambda m_\tau\,d^{(-\kappa)(-\lambda)(-\tau)}=0\,.
\eeq
Extracting the term $m_{-\kappa} m_{-\lambda} m_{-\tau}$ of this relation, we find
\beq
t^\kappa t^\lambda t^\tau+{\rm perms}=\frac{1}{2}\Big(\delta^{\kappa(-\lambda)}t^\tau+\delta^{\lambda(-\tau)}t^\kappa+\delta^{\tau(-\kappa)}t^\lambda\Big)+\frac{1}{2}d^{\kappa\lambda\tau}\,.\label{eq:44}
\eeq
The left hand-side writes $\{\{t^\kappa,t^\lambda\},t^\tau\}+t^\kappa t^\tau t^\lambda+t^\lambda t^\tau t^\kappa$. If we symmetrize Eq.~(\ref{eq:44}) with respect to $\kappa$ and $\tau$, we find
\beq
\{\{t^\kappa, t^\lambda\},t^\tau\}+\{\{t^\lambda, t^\tau\},t^\kappa\}+\{\{t^\tau, t^\kappa\},t^\lambda\}=\delta^{\kappa(-\lambda)}t^\tau+\delta^{\lambda(-\tau)}t^\kappa+\delta^{\tau(-\kappa)}t^\lambda+d^{\kappa\lambda\tau}.
\eeq
Finally, we have
\beq
\{\{t^\kappa, t^\lambda\},t^\tau\}=\frac{2}{3}\delta^{\kappa(-\lambda)}t^\tau+\frac{1}{3}d^{\kappa\lambda\tau}+d^{\kappa\lambda(-\eta)}d^{\tau\sigma\eta}t^{-\sigma}\,.
\eeq
Plugging this in the previous expression, we end up with
\beq
d^{\kappa\lambda(-\eta)}d^{\tau\sigma\eta}+d^{\lambda\tau(-\eta)}d^{\kappa\sigma\eta}+d^{\tau\kappa(-\eta)}d^{\lambda\sigma\eta}=\frac{1}{3}\Big(\delta^{\kappa(-\lambda)}\delta^{\tau(-\sigma)}+\delta^{\lambda(-\tau)}\delta^{\kappa(-\sigma)}+\delta^{\tau(-\kappa)}\delta^{\lambda(-\sigma)}\Big).\nonumber\\
\eeq
Combining this identity with (\ref{eq:KK}) in which we have permuted the indices $\kappa$, $\lambda$ and $\tau$
\beq
f^{\lambda\tau(-\eta)}f^{\kappa\sigma\eta} & = & -\left[\frac{2}{3}\Big(\delta^{\lambda(-\kappa)}\delta^{\tau(-\sigma)}-\delta^{\lambda(-\sigma)}\delta^{\tau(-\kappa)}\Big)+d^{\lambda\kappa(-\eta)}d^{\tau\sigma\eta}-d^{\lambda\sigma(-\eta)}d^{\tau\kappa\eta}\right],\\
f^{\tau\kappa(-\eta)}f^{\lambda\sigma\eta} & = & -\left[\frac{2}{3}\Big(\delta^{\tau(-\lambda)}\delta^{\kappa(-\sigma)}-\delta^{\tau(-\sigma)}\delta^{\kappa(-\lambda)}\Big)+d^{\tau\lambda(-\eta)}d^{\kappa\sigma\eta}-d^{\tau\sigma(-\eta)}d^{\kappa\lambda\eta}\right],
\eeq
we find\footnote{Again, this should be compared to a similar formula obtained using Cartesian color bases \cite{su3}.}
\beq
3d^{\kappa\lambda(-\eta)}d^{\tau\sigma\eta} & = & \delta^{\kappa(-\tau)}\delta^{\lambda(-\sigma)}+\delta^{\kappa(-\sigma)}\delta^{\lambda(-\tau)}-\delta^{\kappa(-\lambda)}\delta^{\tau(-\sigma)}-\Big(f^{\kappa\tau(-\eta)}f^{\lambda\sigma\eta}+f^{\kappa\sigma(-\eta)}f^{\lambda\tau\eta}\Big).\nonumber\\
\eeq
In particular
\beq
3\sum_\tau |d^{\kappa\lambda\tau}|^2 & = & 3\,d^{\kappa\lambda(-\eta)}d^{(-\kappa)(-\lambda)\eta}\nonumber\\\
 & = & 1+\delta^{\kappa\lambda}-\delta^{\kappa(-\lambda)}-f^{\kappa(-\kappa)(-\eta)}f^{\lambda(-\lambda)\eta}-\sum_\tau |f^{\kappa(-\lambda)\tau}|^2\,.
 \label{lbr}
\eeq
The Jacobi identity gives
\beq
0 & = & f^{\kappa(-\kappa)(-\eta)}f^{\lambda(-\lambda)\eta}+f^{(-\kappa)\lambda(-\eta)}f^{\kappa(-\lambda)\eta}+f^{\lambda\kappa(-\eta)}f^{(-\kappa)(-\lambda)\eta}\nonumber\\
& = & f^{\kappa(-\kappa)(-\eta)}f^{\lambda(-\lambda)\eta} -\sum_\tau |f^{\kappa(-\lambda)\tau}|^2+\sum_\tau |f^{\kappa\lambda\tau}|^2
\eeq
and thus
\beq
3\sum_\tau |d^{\kappa\lambda\tau}|^2=1+\delta^{\kappa\lambda}-\delta^{\kappa(-\lambda)}-2f^{\kappa(-\kappa)(-\eta)}f^{\lambda(-\lambda)\eta}-\sum_\tau |f^{\kappa\lambda\tau}|^2\,.
\eeq
We shall later show that $f^{\kappa(-\kappa)\alpha}=0$ and $f^{\kappa(-\kappa)0^{(j)}}=\kappa^j$. It follows then that
\beq
3\sum_\tau |d^{\kappa\lambda\tau}|^2 & = & 1+\delta^{\kappa\lambda}-\delta^{\kappa(-\lambda)}-2(\kappa\cdot\lambda)-\sum_\tau |f^{\kappa\lambda\tau}|^2\,.
\eeq

\section{Cyclic traces}\label{sec:traces}
We are now ready to evaluate traces of products of various $f$'s and $d$'s. We shall write them generally as
\beq
{\cal T}^{\lambda_1\cdots\lambda_n}_{e_1\cdots e_n}\equiv e_1^{\kappa_1\lambda_1(-\kappa_2)}e_2^{\kappa_2\lambda_2(-\kappa_3)}\cdots e_n^{\kappa_n\lambda_n(-\kappa_1)}
\eeq
with each $e_i$ standing for $d$ or $f$ and where a summation over the labels $\kappa_i$ is implied. We mention that ${\cal T}^{\lambda_1\cdots\lambda_n}_{e_1\cdots e_n}$ is invariant under simultaneous cyclic permutations of the indices $e_i$ and $\lambda_i$:
\beq\label{eq:C_cycl}
{\cal T}^{\lambda_1\lambda_2\cdots\lambda_n}_{e_1e_2\cdots e_n}={\cal T}^{\lambda_2\cdots\lambda_n\lambda_1}_{e_2\cdots e_ne_1}\,.
\eeq 
Moreover, from the symmetry of $d$ and the anti-symmetry of $f$, it follows that
\beq\label{eq:C_tr}
{\cal T}^{\lambda_1\cdots\lambda_n}_{e_1\cdots e_n}=(-1)^{\#_f}{\cal T}^{\lambda_n\cdots\lambda_2\lambda_1}_{e_n\cdots e_2e_1}\,,
\eeq 
where $\#_f$ is the number of occurrences of $f$ among the $e_i$'s. Finally, from Eq.~(\ref{eq:cc}), one deduces that
\beq
({\cal T}^{\lambda_1\cdots\lambda_n}_{e_1\cdots e_n})^*=(-1)^{\#_f}{\cal T}^{(-\lambda_1)\cdots(-\lambda_n)}_{e_1\cdots e_n}\,.
\eeq
In a few occasions, we shall also need the matrix
\beq
({\cal M}^{\lambda_1\cdots\lambda_n}_{e_1\cdots e_n})^{\kappa\tau}\equiv e_1^{\kappa\lambda_1(-\kappa_2)}e_2^{\kappa_2\lambda_2(-\kappa_3)}\cdots e_n^{\kappa_n\lambda_n(-\tau)}\,,
\eeq
whose trace is ${\cal T}^{\lambda_1\cdots\lambda_n}_{e_1\cdots e_n}$.

\subsection{Color traces with $n=1$}
Since $t^{-\kappa}t^\kappa$ is proportional to the identity and ${\rm tr} \,t^{\kappa}=0$, we can see from eq. \eqref{eq:C9} that
\beq
{\cal T}^\lambda_d={\cal T}^\lambda_f=0\,.
\eeq

\subsection{Color traces with $n=2$}
From the symmetry of $d$ and the anti-symmetry of $f$, we have ${\cal T}_{fd}^{\lambda\lambda'}\equiv f^{\kappa\lambda(-\tau)}d^{\tau\lambda'(-\kappa)}=f^{\kappa\lambda(-\tau)}(d^{(-\tau)(-\lambda')\kappa})^*=0$.\footnote{For the same reason, we have $f^{\kappa\lambda\tau}d^{\tau\lambda'\kappa}=0$.} To determine ${\cal T}_{ff}^{\lambda\lambda'}\equiv f^{\kappa\lambda(-\tau)}f^{\tau\lambda'(-\kappa)}$, we use Eq.~(\ref{eq:C9}) and write
\beq\label{eq:FF}
{\cal T}_{ff}^{\lambda\lambda'}=4\,({\rm tr}\,t^\kappa t^\lambda t^{-\tau}-{\rm tr}\,t^\lambda t^\kappa t^{-\tau})({\rm tr}\,t^\tau t^{\lambda'} t^{-\kappa}-{\rm tr}\,t^{\lambda'} t^\tau t^{-\kappa})\,.
\eeq
This expression can be simplified by using the Fierz identity (\ref{eq:Fierz}) repeatedly. For instance, one writes
\beq
{\rm tr}\,t^\kappa t^\lambda t^{-\tau}\, {\rm tr}\,t^\tau t^{\lambda'} t^{-\kappa} & = & \frac{1}{2}{\rm tr}\,t^\kappa t^\lambda t^{\lambda'} t^{-\kappa}-\frac{1}{2N}{\rm tr}\,t^\kappa t^\lambda\, {\rm tr}\,t^{\lambda'} t^{-\kappa}\nonumber\\
& = & \frac{N^2-1}{4N}{\rm tr}\,t^\lambda t^{\lambda'}-\frac{1}{2N}{\rm tr}\,t^\kappa t^\lambda\, {\rm tr}\,t^{\lambda'} t^{-\kappa}\nonumber\\
& = & \frac{N^2-1}{8N}\delta^{\lambda(-\lambda')}-\frac{1}{8N}\delta^{\lambda(-\kappa)}\delta^{(-\kappa)(-\lambda')}=\frac{N^2-2}{8N}\delta^{\lambda(-\lambda')}\,.
\eeq
Similarly,
\beq
{\rm tr}\,t^\kappa t^\lambda t^{-\tau}\, {\rm tr}\,t^{\lambda'}t^\tau  t^{-\kappa} & = & \frac{1}{2}{\rm tr}\,t^\kappa t^\lambda t^{-\kappa}t^{\lambda'} -\frac{1}{2N}{\rm tr}\,t^\kappa t^\lambda\, {\rm tr}\,t^{\lambda'} t^{-\kappa}\nonumber\\
& = & -\frac{1}{4N}{\rm tr}\,t^\lambda t^{\lambda'}-\frac{1}{2N}{\rm tr}\,t^\kappa t^\lambda\, {\rm tr}\,t^{\lambda'} t^{-\kappa}=-\frac{1}{4N}\delta^{\lambda(-\lambda')}\,.
\eeq
Combining these and other similar contributions into Eq.~(\ref{eq:FF}), one finds
\beq
{\cal T}_{ff}^{\lambda\lambda'}=4\left(\frac{N^2-2}{8N}+\frac{N^2-2}{8N}+\frac{1}{4N}+\frac{1}{4N}\right)\delta^{\lambda(-\lambda')}=N\delta^{\lambda(-\lambda')}\,.\label{eq:65}
\eeq
Similarly
\beq
{\cal T}_{dd}^{\lambda\lambda'}=4\left(\frac{N^2-2}{8N}+\frac{N^2-2}{8N}-\frac{1}{4N}-\frac{1}{4N}\right)\delta^{\lambda(-\lambda')}=\frac{N^2-4}{N}\delta^{\lambda(-\lambda')}\,.\label{eq:66}
\eeq
This type of calculations can be conveniently organized and performed by associating a diagrammatic representation to the Fierz identity as well as to Eq.~(\ref{eq:C9}). Summarizing our results for $n=2$, we have found that
\beq\label{eq:HH}
{\cal T}^{\lambda\lambda'}_{ff}=N\delta^{\lambda(-\lambda')}\,,\quad {\cal T}^{\lambda\lambda'}_{dd}=\frac{N^2-4}{N}\delta^{\lambda(-\lambda')}\,,\quad {\cal T}^{\lambda\lambda'}_{df}={\cal T}^{\lambda\lambda'}_{fd}=0\,.
\eeq

\subsection{Color traces with $n=3$}
We could continue using the Fierz identity. However, it is more convenient here to use the Jacobi identities (to which one can also associate diagrammatic rules). In fact, the Jacobi identities can be turned into useful identities among the various color traces. From Eqs.~(\ref{eq:GG})-(\ref{eq:II}), one finds
\beq
{\cal T}^{\lambda\lambda'\lambda_1\cdots\lambda_n}_{ffe_1\cdots e_n} & = & f^{\lambda\lambda'(-\eta)}{\cal T}^{\eta\lambda_1\cdots\lambda_n}_{fe_1\cdots e_n}+{\cal T}^{\lambda'\lambda\lambda_1\cdots\lambda_n}_{ffe_1\cdots e_n}\,,\label{eq:C_one}\\
{\cal T}^{\lambda\lambda'\lambda_1\cdots\lambda_n}_{dfe_1\cdots e_n} & = & d^{\lambda\lambda'(-\eta)}{\cal T}^{\eta\lambda_1\cdots\lambda_n}_{fe_1\cdots e_n}-{\cal T}^{\lambda'\lambda\lambda_1\cdots\lambda_n}_{dfe_1\cdots e_n}\,,\label{eq:C_two}\\
{\cal T}^{\lambda\lambda'\lambda_1\cdots\lambda_n}_{dfe_1\cdots e_n} & = & f^{\lambda\lambda'(-\eta)}{\cal T}^{\eta\lambda_1\cdots\lambda_n}_{de_1\cdots e_n}+{\cal T}^{\lambda'\lambda\lambda_1\cdots\lambda_n}_{fde_1\cdots e_n}\,.\label{eq:C_three}
\eeq
From Eq.~(\ref{eq:KK}), we also find
\beq
{\cal T}^{\lambda\lambda'\lambda_1\cdots\lambda_n}_{ffe_1\cdots e_n}=\frac{2}{N}\Big(\delta^{\lambda(-\lambda')}{\cal T}^{\lambda_1\cdots\lambda_n}_{e_1\cdots e_n}-({\cal M}^{\lambda_1\cdots\lambda_n}_{e_1\cdots e_n})^{\lambda(-\lambda')}\Big)+d^{\lambda\lambda'(-\eta)}{\cal T}^{\eta\lambda_1\cdots\lambda_n}_{de_1\cdots e_n}-{\cal T}^{\lambda'\lambda\lambda_1\cdots\lambda_n}_{dde_1\cdots e_n}\,.\label{eq:C_four}
\eeq
Using  Eqs.~(\ref{eq:C_one}), (\ref{eq:HH}) and (\ref{eq:C_cycl})-(\ref{eq:C_tr}), we find
\beq
{\cal T}^{\kappa\lambda\tau}_{fff} & = & f^{\kappa\lambda(-\eta)}{\cal T}^{\eta\tau}_{ff}+{\cal T}^{\lambda\kappa\tau}_{fff}\nonumber\\
& = & f^{\kappa\lambda(-\eta)}{\cal T}^{\eta\tau}_{ff}-{\cal T}^{\kappa\lambda\tau}_{fff}\nonumber\\
& = & \frac{1}{2}f^{\kappa\lambda(-\eta)}{\cal T}^{\eta\tau}_{ff}=\frac{N}{2}f^{\kappa\lambda\tau}\,.
\eeq
Similarly, using Eqs.~(\ref{eq:C_three}), (\ref{eq:HH}) and (\ref{eq:C_cycl})-(\ref{eq:C_tr}), we find
\beq
{\cal T}^{\kappa\lambda\tau}_{dfd} & = & f^{\kappa\lambda(-\eta)}{\cal T}^{\eta\tau}_{dd}+{\cal T}^{\lambda\kappa\tau}_{fdd}\nonumber\\
& = & f^{\kappa\lambda(-\eta)}{\cal T}^{\eta\tau}_{dd}-{\cal T}^{\kappa\lambda\tau}_{dfd}\nonumber\\
& = & \frac{1}{2}f^{\kappa\lambda(-\eta)}{\cal T}^{\eta\tau}_{dd}=\frac{N^2-4}{2N}f^{\kappa\lambda\tau}\,.
\eeq
And using Eqs.~(\ref{eq:C_two}), (\ref{eq:HH}) and (\ref{eq:C_cycl})-(\ref{eq:C_tr}), we find
\beq
{\cal T}^{\kappa\lambda\tau}_{dff} & = & d^{\kappa\lambda(-\eta)}{\cal T}^{\eta\tau}_{ff}-{\cal T}^{\lambda\kappa\tau}_{dff}\nonumber\\
& = & d^{\kappa\lambda(-\eta)}{\cal T}^{\eta\tau}_{ff}-d^{\lambda\tau(-\eta)}{\cal T}^{\eta\kappa}_{ff}+{\cal T}^{\tau\lambda\kappa}_{dff}\nonumber\\
& = & d^{\kappa\lambda(-\eta)}{\cal T}^{\eta\tau}_{ff}-d^{\lambda\tau(-\eta)}{\cal T}^{\eta\kappa}_{ff}+d^{\tau\kappa(-\eta)}{\cal T}^{\eta\lambda}_{ff}-{\cal T}^{\kappa\lambda\tau}_{dff}\nonumber\\
& = & \frac{1}{2}\Big(d^{\kappa\lambda(-\eta)}{\cal T}^{\eta\tau}_{ff}-d^{\lambda\tau(-\eta)}{\cal T}^{\eta\kappa}_{ff}+d^{\tau\kappa(-\eta)}{\cal T}^{\eta\lambda}_{ff}\Big)=\frac{1}{2}d^{\kappa\lambda(-\eta)}{\cal T}^{\eta\tau}_{ff}=\frac{N}{2}d^{\kappa\lambda\tau}\,.
\eeq
From Eq.~(\ref{eq:C_four}), we have
\beq
{\cal T}^{\kappa\lambda\tau}_{ffd}=\frac{2}{N}\Big(\delta^{\kappa(-\lambda)}{\cal T}^{\tau}_d-({\cal M}^{\tau}_d)^{\lambda(-\kappa)}\Big)+d^{\kappa\lambda(-\eta)}{\cal T}^{\eta\tau}_{dd}-{\cal T}^{\kappa\lambda\tau}_{ddd}\,,
\eeq
from which we deduce that
\beq
{\cal T}^{\kappa\lambda\tau}_{ddd} & = & \frac{2}{N}\Big(\delta^{\kappa(-\lambda)}{\cal T}^{\tau}_d-({\cal M}^{\tau}_d)^{\lambda(-\kappa)}\Big)+d^{\kappa\lambda(-\eta)}{\cal T}^{\eta\tau}_{dd}-{\cal T}^{\kappa\lambda\tau}_{ffd}\nonumber\\
& = & \left(-\frac{2}{N}+\frac{N^2-4}{N}-\frac{N}{2}\right)d^{\kappa\lambda\tau}=\frac{N^2-12}{2N}d^{\kappa\lambda\tau}\,.
\eeq
Summarizing our results for $n=3$, we have found that
\beq
{\cal T}^{\kappa\lambda\tau}_{fff}=\frac{N}{2}f^{\kappa\lambda\tau}\!, \,\,\,\, {\cal T}^{\kappa\lambda\tau}_{dff}=\frac{N}{2}d^{\kappa\lambda\tau}\!, \,\,\,\, {\cal T}^{\kappa\lambda\tau}_{ddf}=\frac{N^2-4}{2N}f^{\kappa\lambda\tau}\!, \,\,\,\, {\cal T}^{\kappa\lambda\tau}_{ddd}=\frac{N^2-12}{2N}d^{\kappa\lambda\tau}.\label{eq:fff}
\eeq
As already mentioned, these results can also be obtained using the Fierz identity and the associated diagrammatic rules.

\subsection{Color traces with $n=4$}
Here we discuss only ${\cal T}^{\kappa\lambda\tau\sigma}_{ffff}$. We can proceed in two different ways. Using the Fierz identity and the associated diagrammatic rules (calculation not detailed here), we find
\beq
{\cal T}^{\kappa\lambda\tau\sigma}_{ffff} & = & \frac{1}{2}\Big(\delta^{\kappa(-\lambda)}\delta^{\tau(-\sigma)}+\delta^{\kappa(-\tau)}\delta^{\lambda(-\sigma)}+\delta^{\kappa(-\sigma)}\delta^{\lambda(-\tau)}\Big)\nonumber\\
& + & N\,{\rm tr}\,\Big(t^\kappa t^\lambda t^\tau t^\sigma+t^\lambda t^\kappa t^\sigma t^\tau\Big).
\eeq
This expression makes the symmetries of ${\cal T}^{\kappa\lambda\tau\sigma}_{ffff}$ explicit. Using Eq.~(\ref{eq:CC}), we arrive at\footnote{The corresponding identity withing a Cartesian basis writes
\beq
{\cal T}^{abcd}_{ffff}=\delta^{ab}\delta^{cd}+\frac{1}{2}\Big(\delta^{ac}\delta^{bd}+\delta^{ad}\delta^{bc}\Big)+\frac{N}{4}\Big(d^{abe} d^{cde}-f^{abe}f^{cde}\Big).\nonumber
\eeq}
\beq\label{eq:ffff}
{\cal T}^{\kappa\lambda\tau\sigma}_{ffff} & = & \delta^{\kappa(-\lambda)}\delta^{\tau(-\sigma)}+\frac{1}{2}\Big(\delta^{\kappa(-\tau)}\delta^{\lambda(-\sigma)}+\delta^{\kappa(-\sigma)}\delta^{\lambda(-\tau)}\Big)\nonumber\\
& + & \frac{N}{4}\Big(d^{\kappa\lambda\eta} d^{\tau\sigma(-\eta)}+f^{\kappa\lambda\eta}f^{\tau\sigma(-\eta)}\Big).
\eeq
Let us now re-derive this result using the Jacobi identities. From Eq.~(\ref{eq:C_four}), we find
\beq\label{eq:idk}
{\cal T}^{\kappa\lambda\tau\sigma}_{ffff} & = & \frac{2}{N}\Big(\delta^{\kappa(-\lambda)}{\cal T}^{\tau\sigma}_{ff}-({\cal M}^{\tau\sigma}_{ff})^{\kappa(-\lambda)}\Big)+d^{\kappa\lambda(-\eta)}{\cal T}^{\eta\tau\sigma}_{dff}-{\cal T}^{\lambda\kappa\tau\sigma}_{ddff}\nonumber\\
& = & 2\delta^{\kappa(-\lambda)}\delta^{\tau(-\sigma)}-\frac{2}{N}f^{\kappa\tau(-\eta)}f^{\eta\sigma\lambda}+\frac{N}{2}d^{\kappa\lambda(-\eta)}d^{\eta\tau\sigma}-{\cal T}^{\lambda\kappa\tau\sigma}_{ddff}\,.
\eeq
From Eqs.~(\ref{eq:C_one}) and (\ref{eq:C_two}), we have
\beq
{\cal T}^{\lambda\kappa\tau\sigma}_{ddff} & = & f^{\tau\sigma(-\eta)}{\cal T}^{\eta\lambda\kappa}_{fdd}+{\cal T}^{\lambda\kappa\sigma\tau}_{ddff}\nonumber\\
& = & f^{\tau\sigma(-\eta)}{\cal T}^{\eta\lambda\kappa}_{fdd}+d^{\kappa\sigma(-\eta)}{\cal T}^{\eta\tau\lambda}_{ffd}-{\cal T}^{\lambda\sigma\kappa\tau}_{ddff}\,,\label{eq:81}
\eeq
which gives ${\cal T}^{\lambda\kappa\tau\sigma}_{ddff}+{\cal T}^{\lambda\sigma\kappa\tau}_{ddff}$ in terms of known quantities. This quantity is not the one that appears in (\ref{eq:idk}), but rather in ${\cal T}^{\kappa\lambda\tau\sigma}_{ffff}+{\cal T}^{\sigma\lambda\kappa\tau}_{ffff}$. However, using the previously derived identities, it is easily seen that
\beq
\frac{1}{2}\Big({\cal T}^{\kappa\lambda\tau\sigma}_{ffff}+{\cal T}^{\sigma\lambda\kappa\tau}_{ffff}\Big)+\frac{N}{4}f^{\kappa\lambda(-\eta)}f^{\tau\sigma\eta}={\cal T}^{\kappa\lambda\tau\sigma}_{ffff}\,.
\eeq
By plugging Eq.~(\ref{eq:idk}) in the left-hand side and upon using (\ref{eq:81}), we find
\beq
{\cal T}^{\kappa\lambda\tau\sigma}_{ffff} & = & \delta^{\kappa(-\lambda)}\delta^{\tau(-\sigma)}+\delta^{\sigma(-\lambda)}\delta^{\kappa(-\tau)}-\frac{1}{N}\Big(f^{\kappa\tau(-\eta)}f^{\eta\sigma\lambda}+f^{\sigma\kappa(-\eta)}f^{\eta\tau\lambda}\Big)\nonumber\\
& + & \frac{N}{4}\Big(d^{\kappa\lambda(-\eta)}d^{\eta\tau\sigma}+d^{\sigma\lambda(-\eta)}d^{\eta\kappa\tau}\Big)-\frac{1}{2}\Big({\cal T}^{\lambda\kappa\tau\sigma}_{ddff}+{\cal T}^{\lambda\sigma\kappa\tau}_{ddff}\Big)+\frac{N}{4}f^{\kappa\lambda(-\eta)}f^{\tau\sigma\eta}\nonumber\\
& = & \delta^{\kappa(-\lambda)}\delta^{\tau(-\sigma)}+\delta^{\sigma(-\lambda)}\delta^{\kappa(-\tau)}-\frac{1}{N}\Big(f^{\kappa\tau(-\eta)}f^{\eta\sigma\lambda}+f^{\sigma\kappa(-\eta)}f^{\eta\tau\lambda}+f^{\kappa\lambda(-\eta)}f^{\tau\sigma\eta}\Big)\nonumber\\
& + & \frac{N}{4}\Big(d^{\kappa\lambda(-\eta)}d^{\eta\tau\sigma}+d^{\sigma\lambda(-\eta)}d^{\eta\kappa\tau}-d^{\kappa\sigma(-\eta)}d^{\eta\tau\lambda}\Big)+\frac{N}{2}f^{\kappa\lambda(-\eta)}f^{\tau\sigma\eta}\,,
\eeq 
where we used Eq.~(\ref{eq:fff}). Using Eqs.~(\ref{eq:GG}) and (\ref{eq:KK}), we arrive at the announced result (\ref{eq:ffff}).

\section{Summing over zeros}\label{sec:zeros}
As recalled in the Introduction, in the presence  of constant temporal backgrounds, the color indices are not restricted to the structure constants, but appear generically within the Feynman integrals via the shift of the momenta $P_\mu\to P_\mu^\kappa=P_\mu+T(r\cdot\kappa) \delta_{\mu0}$. A direct consequence of this is that the summation over all internal color indices of a given diagram cannot be explicitly performed. However, since the momenta are not shifted for color modes corresponding to zeros (since $r\cdot\kappa=0$ in this case), one can usually perform a partial summation over the color indices. Here, without aiming at full generality, we illustrate this partial summation over zeros on  sums of the form
\beq
{\cal F}[X]\equiv\sum_{\kappa\lambda\tau} |f^{\kappa\lambda\tau}|^2 X^{\kappa\lambda\tau} \quad \mbox{and} \quad {\cal D}[X]\equiv\sum_{\kappa\lambda\tau} |d^{\kappa\lambda\tau}|^2 X^{\kappa\lambda\tau}\,.\label{eq:sums}
\eeq
The first type of sum appears for instance in the evaluation of the two-loop background effective potential in the Faddeev-Popov and Curci-Ferrari models \cite{Reinosa:2015gxn}. The second one appears in the evaluation of the two-loop background effective potential in the Gribov-Zwanziger inspired model of Ref.~\cite{Kroff:2018ncl}, see Ref.~\cite{second} for more details.

In the absence of background, $X^{\kappa\lambda\tau}$ does not depend on the color labels and the sums (\ref{eq:sums}) boil down to the evaluation of $\sum_{\kappa\lambda\tau} |f^{\kappa\lambda\tau}|^2$ and $\sum_{\kappa\lambda\tau} |d^{\kappa\lambda\tau}|^2$ which are easily found to be given by
\beq
{\cal F}[1]\equiv\sum_{\kappa\lambda\tau} |f^{\kappa\lambda\tau}|^2=N(N^2-1)\,, \quad {\cal D}[1]\equiv\sum_{\kappa\lambda\tau} |d^{\kappa\lambda\tau}|^2=\frac{(N^2-1)(N^2-4)}{N}\,, 
\eeq
from Eqs.~(\ref{eq:65})-(\ref{eq:66}) by setting $\lambda'=-\lambda$ and summing over $\lambda$ (these are of course the standard results that one would obtain by using the standard color bases). We now explain how to extend these formulas to the case where $X^{\kappa\lambda\tau}$ depends on the color indices. The key property will be the evaluation (and the geometrical interpretation) of the various tensor components $f^{\kappa\lambda\tau}$ and $d^{\kappa\lambda\tau}$, which we discuss in the next section.

\subsection{Values of the tensors $f^{\kappa\lambda\tau}$ and $d^{\kappa\lambda\tau}$}\label{sec:values}
Let us first derive the following property: $f^{\kappa\lambda\tau}=d^{\kappa\lambda\tau}=0$ if $\kappa+\lambda+\tau\neq 0$. To see this, let us evaluate $[t^{0^{(j)}},t^\kappa t^\lambda]$. We find
\beq
[t^{0^{(j)}},t^\kappa t^\lambda] & = & [t^{0^{(j)}},t^\kappa]t^\lambda+t^\kappa[t^{0^{(j)}}, t^\lambda]\nonumber\\
& = & \kappa^j\,t^\kappa t^\lambda+\lambda^j\,t^\kappa t^\lambda=(\kappa+\lambda)^j\,t^\kappa t^\lambda\,.
\eeq
This implies that the component of $t^\kappa t^\lambda$ that belongs to the algebra, that is the second term in Eq.~(\ref{eq:CC}), lies in the eigenspace with adjoint weight $\kappa+\lambda$, that is it needs to be along the generator $t^{-\tau}$ with $-\tau=\kappa+\lambda$. It follows that the structure constants do not have any contribution if $\kappa+\lambda+\tau\neq 0$, as announced.\footnote{In particular, this means that they are zero if $\kappa+\lambda$ is not an adjoint weight.} 

This leaves us with three cases where $f^{\kappa\lambda\tau}$ and $d^{\kappa\lambda\tau}$ can have non-zero values: 1) $\kappa$, $\lambda$ and $\tau$ are roots $\alpha$, $\beta$ and $\gamma$ such that $\alpha+\beta+\gamma=0$; 2) one of the color labels is a zero and the other two are opposite roots; 3) the three color labels are zeros (in this case only $d^{\kappa\lambda\tau}$ can have non-zero values since the second equation in (\ref{eq:C9}) gives obviously $0$). We shall now evaluate the structure constants in these three cases separately.\\

Let us first consider two generators of the form $t^\alpha$ and $t^\beta$, which we rewrite as $Z^{jj'}$ and $Z^{ll'}$ with $\alpha=\rho_j-\rho_{j'}$ and $\beta=\rho_l-\rho_{l'}$. Multiplying the two matrices, one easily finds $Z^{jj'}Z^{ll'}=Z^{jl'}\delta_{j'l}/\sqrt{2}$ and thus
\beq
{[}Z^{jj'},Z^{ll'}{]}=\frac{Z^{jl'}\delta_{j'l}-Z^{lj'}\delta_{l'j}}{\sqrt{2}}\,, \quad \{Z^{jj'},Z^{ll'}\}=\frac{Z^{jl'}\delta_{j'l}+Z^{lj'}\delta_{l'j}}{\sqrt{2}}\,.\label{eq:truc}
\eeq
To obtain non-vanishing structure constants, we need to take $j'=l$ or $j=l'$, or both. Taking $j'=l$ (but $j\neq l'$), one finds
\beq
[Z^{jj'},Z^{ll'}]=\{Z^{jj'},Z^{ll'}\}=\frac{1}{\sqrt{2}}Z^{jl'}\,,
\eeq
which is nothing but
\beq
[t^\alpha,t^\beta]=\{t^\alpha,t^\beta\}=\frac{1}{\sqrt{2}}t^{-\gamma}\,,
\eeq
with $\alpha=\rho^\alpha-\bar\rho^\alpha$, $\beta=\rho^\beta-\bar\rho^\beta$ and $\gamma=\rho^\gamma-\bar\rho^\gamma$, with $\rho^\beta=\bar\rho^\alpha$, $\rho^\gamma=\bar\rho^\beta$ and $\rho^\alpha=\bar\rho^\gamma$. Similarly, taking $j=l'$ (but $j'\neq l$), one finds
\beq
[Z^{jj'},Z^{ll'}]=-\{Z^{jj'},Z^{ll'}\}=-\frac{1}{\sqrt{2}}Z^{lj'}\,,
\eeq
which is nothing but
\beq
[t^\alpha,t^\beta]=-\{t^\alpha,t^\beta\}=-\frac{1}{\sqrt{2}}t^{-\gamma}\,,
\eeq
with $\alpha=\rho^\alpha-\bar\rho^\alpha$, $\beta=\rho^\beta-\bar\rho^\beta$ and $\gamma=\rho^\gamma-\bar\rho^\gamma$, with $\bar\rho^\beta=\rho^\alpha$, $\bar\rho^\gamma=\rho^\beta$ and $\bar\rho^\alpha=\rho^\gamma$. We show in App.~\ref{app:sum} that these are the only two possible situations where three roots $\alpha$, $\beta$ and $\gamma$ can sum to zero, and, therefore, we can write
\beq
f^{\alpha\beta\gamma}=\sigma d^{\alpha\beta\gamma}=\sigma \frac{\delta_{\alpha+\beta+\gamma,0}}{\sqrt{2}}\,,\label{eq:90}
\eeq
where $\sigma=+1$ if $\rho^\beta=\bar\rho^\alpha$ and $\sigma=-1$ if $\bar\rho^\beta=\rho^\alpha$.

We still need to consider the case where both $j=l'$ and $j'=l$ in Eq.~(\ref{eq:truc}), which then becomes
\beq
{[}Z^{jl},Z^{lj}{]}=\frac{Z^{jj}-Z^{ll}}{\sqrt{2}}\,, \quad \{Z^{jl},Z^{lj}\}=\frac{Z^{jj}+Z^{ll}}{\sqrt{2}}\,.
\eeq
Since $Z^{jj}$ an $Z^{ll}$ are not elements of the Cartan-Weyl basis, we need to expand them along $\{H^j,Z^{jj'}\}$. But this is easily done since $Z^{jj}$ is diagonal so it does not have components along the $Z^{jj'}$. Using Eq.~(\ref{eq:C3}), we find
\beq
[Z^{jl},Z^{lj}] & = & \sum_k (H^k_{jj}-H^k_{ll})H^k=\sum_k (\rho^k_j-\rho^k_l)H^k\,,\\
\{Z^{jl},Z^{lj}\} & = & \frac{\mathds{1}}{N}+ \sum_k (H^k_{jj}+H^k_{ll})H^k=\frac{\mathds{1}}{N}+\sum_k (\rho^k_j+\rho^k_l)H^k\,.
\eeq
This is nothing but
\beq
[t^\alpha,t^{-\alpha}] & = & \sum_k (\rho^\alpha-\rho^{\bar\alpha})^kt^{0^{(k)}}\,,\\
\{t^\alpha,t^{-\alpha}\} & = & \frac{\mathds{1}}{N}+\sum_k (\rho^\alpha+\rho^{\bar\alpha})^kt^{0^{(k)}}\,,
\eeq
with $\rho^\alpha$ and $\rho^{\bar\alpha}$ such that $\alpha=\rho^\alpha-\rho^{\bar\alpha}$. It follows that 
\beq
f^{\alpha(-\alpha)0^{(j)}}=(\rho^\alpha-\rho^{\bar\alpha})^j \quad {\rm and} \quad d^{\alpha(-\alpha)0^{(j)}}=(\rho^\alpha+\rho^{\bar\alpha})^j\,.\label{eq:0rmr}
\eeq
This provides a neat geometrical interpretation of the structure constants which one could use as a mnemonic way to remember them.

We finally need to consider the case where the three color labels are zeros (in which case the only non-vanishing components are those of $d^{\kappa\lambda\tau}$). We have $d^{0^{(j)}0^{(k)}0^{(l)}}=4\,{\rm tr}\,H^jH^kH^l$. Suppose that $j$, $k$ and $l$ are all different and assume without loss of generality that $j$ is the smallest. From the block diagonal structure of the $H^j$, it is easily seen that $H^jH^kH^l\propto H^j$ and then $d^{0^{(j)}0^{(k)}0^{(l)}}=0$ in this case. It follows that, in order to obtain non-vanishing structure constants, we need at least two zeros with the same multiplicity index. Without loss of generality, we can treat the three cases $j<k=l$, $j=k<l$ and $j=k=l$. The first case gives $\smash{d^{0^{(j)}0^{(k)}0^{(k)}}=0}$. In the second case, we find
\beq
d^{0^{(k)}0^{(k)}0^{(l)}}=\frac{4(k+k^2)}{2k(k+1)\sqrt{2l(l+1)}}=\sqrt{\frac{2}{l(l+1)}}\,,\label{eq:z1}
\eeq
whereas in the third case we find
\beq
d^{0^{(l)}0^{(l)}0^{(l)}}=\frac{4(l-l^3)}{2l(l+1)\sqrt{2l(l+1)}}=\sqrt{\frac{2}{l(l+1)}}(1-l)\,.\label{eq:z2}
\eeq

\subsection{Summation formulas}
We have now all the needed properties to evaluate the sums $\sum_{\kappa\lambda\tau} |f^{\kappa\lambda\tau}|^2X^{\kappa\lambda\tau}$, $\sum_{\kappa\lambda\tau} |d^{\kappa\lambda\tau}|^2X^{\kappa\lambda\tau}$. First, we write
\beq
{\cal F}[X] & = & \sum_{\alpha,\,j} |f^{0^{(j)}\alpha(-\alpha)}|^2\Big(X^{0\alpha(-\alpha)}+X^{\alpha 0 (-\alpha)}+X^{\alpha(-\alpha)0}\Big)+\sum_{\alpha\beta\gamma} |f^{\alpha\beta\gamma}|^2X^{\alpha\beta\gamma}\,,\label{eq:99}
\eeq
and
\beq
{\cal D}[X] & = & \Big(\sum_l |d^{\,0^{(l)}0^{(l)}0^{(l)}}|^2+3\sum_{k<l} |d^{0^{(k)}0^{(k)}0^{(l)}}|^2\Big)X^{000}\nonumber\\
& + & \sum_{\alpha,\,j} |d^{\,0^{(j)}\alpha(-\alpha)}|^2\Big(X^{0\alpha(-\alpha)}+X^{\alpha 0 (-\alpha)}+X^{\alpha(-\alpha)0}\Big)+\sum_{\alpha\beta\gamma} |d^{\alpha\beta\gamma}|^2X^{\alpha\beta\gamma}\,,\label{eq:100}
\eeq
where we assume that $X$ is not sensitive to the labelling of the zeros. We then note from Eq.~(\ref{eq:0rmr}) that
\beq
\sum_j |f^{0^{(j)}\alpha(-\alpha)}|^2 & = & (\rho^\alpha)^2+(\bar\rho^\alpha)^2-2\rho^\alpha\cdot\bar\rho^\alpha\,,\\
\sum_j |d^{\,0^{(j)}\alpha(-\alpha)}|^2 & = & (\rho^\alpha)^2+(\bar\rho^\alpha)^2+2\rho^\alpha\cdot\bar\rho^\alpha\,.
\eeq
Using the constraints (\ref{eq:constraints}), we arrive at
\beq
\sum_j |f^{0^{(j)}\alpha(-\alpha)}|^2=1\,, \quad \sum_j |d^{\,0^{(j)}\alpha(-\alpha)}|^2=1-\frac{2}{N}\,.
\eeq
Similarly, using Eqs.~(\ref{eq:z1})-(\ref{eq:z2}), we can evaluate the sum
\beq
& & \sum_l |d^{\,0^{(l)}0^{(l)}0^{(l)}}|^2+3\sum_{k<l} |d^{\,0^{(k)}0^{(k)}0^{(l)}}|^2\nonumber\\
& & \hspace{0.5cm}=\,2\sum_l \frac{(l-1)^2}{l(l+1)}+6\sum_{k<l} \frac{1}{l(l+1)}\nonumber\\
& & \hspace{0.5cm}=\,2\sum_{l=1}^{N-1} \frac{(l-1)(l+2)}{l(l+1)}=2\frac{(N-1)(N-2)}{N}\,.
\eeq
Finally, in Eqs.~(\ref{eq:99})-(\ref{eq:100}), we can restrict the sums that involve three roots $\alpha$, $\beta$ and $\gamma$ to those sets that sum to zero (see App.~\ref{app:sum}), in which case $|f^{\alpha\beta\gamma}|^2=|d^{\alpha\beta\gamma}|^2=1/2$ from Eq.~(\ref{eq:90}). 

Putting all the pieces together, we eventually arrive at
\beq
{\cal F}[X] & = & \sum_{\alpha} \Big(X^{0\alpha(-\alpha)}+X^{\alpha 0 (-\alpha)}+X^{\alpha(-\alpha)0}\Big)+\frac{1}{2}\sum_{(\alpha,\beta,\gamma)} \Big(X^{\alpha\beta\gamma}+{\rm perms.}\Big)\,,
\eeq
and
\beq
{\cal D}[X] & = & 2\frac{(N-1)(N-2)}{N}X^{000}\nonumber\\
& + & \sum_{\alpha} \left(1-\frac{2}{N}\right)\Big(X^{0\alpha(-\alpha)}+X^{\alpha 0 (-\alpha)}+X^{\alpha(-\alpha)0}\Big)+\!\!\frac{1}{2}\sum_{(\alpha,\beta,\gamma)} \Big(X^{\alpha\beta\gamma}+{\rm perms.}\Big)\,,\nonumber\\
\eeq
where $\sum_{(\alpha,\beta,\gamma)}$ sums over the triplets of roots that sum to zero.

In the case where $X$ does not depend on the color indices, we find
\beq
{\cal F}[1] & = & 3\#_\alpha+3\#_{(\alpha,\beta,\gamma)}\,,\\
{\cal D}[1] & = & 2\frac{(N-1)(N-2)}{N}+3\left(1-\frac{2}{N}\right)\#_\alpha+3\#_{(\alpha,\beta,\gamma)}\,,
\eeq
where $\#_\alpha$ and $\#_{(\alpha,\beta,\gamma)}$ denote respectively the number of roots and the number of root triplets that sum to zero. Basic combinatorics leads to
\beq
\#_\alpha=N(N-1)\,, \quad \#_{(\alpha,\beta,\gamma)}=\frac{N(N-1)(N-2)}{3}\,,
\eeq
from which we deduce that
\beq
{\cal F}[1] & = & 3N(N-1)+N(N-1)(N-2)=N(N^2-1)\,,\\
{\cal D}[1] & = & (N-1)(N-2)\left(\frac{2}{N}+3+N\right)=\frac{(N^2-1)(N^2-4)}{N}\,,
\eeq
which are well known results quoted above.

\section{Conclusions}
We have discussed the evaluation of SU($N$) color factors from the point of view of the so-called Cartan-Weyl bases, which are useful in particular in the context of background field methods at finite temperature. One peculiarity of these bases is that the usual Cartesian color indices are replaced by vectors $\kappa$, the so-called adjoint weights, that combine the various commuting charges characterizing a given color mode. Among the many advantages of these adjoint weights is that they greatly clarify the role of the background at finite temperature, the latter appearing as a collection of imaginary chemical potentials associated to the various charges carried by the weight $\kappa$ of a given color mode. As such, the only effect of the background is to shift the frequency of the mode by the amount $T(r\cdot\kappa)\equiv Tr^j\kappa^j$ with $r^j\equiv g\beta\bar A^j$.

This simple interpretation does not come only with advantages since, due to this mixing between color and momentum, the summation over color indices cannot in general be carried out independently from the evaluation of the Feynman integrals. There are some situations, however, where it is possible to do so and, in those cases, the evaluation of color factors is similar to that within Cartesian color bases provided one accounts for the specificities of the Cartan-Weyl bases. We have done so in Secs.~\ref{sec:struc} and \ref{sec:traces} and revisited a certain number of color traces, including both the tensors $f^{\kappa\lambda\tau}$ and $d^{\kappa\lambda\tau}$ (for $N\geq 3$) and even specific formulas in the case $N=3$. In those cases where the weights do not decouple from the momenta, one can still perform explicitly the summations over the zero weights. We have provided formulas for these partial summations in Sec.~\ref{sec:zeros} while giving more details on the tensors $f^{\kappa\lambda\tau}$ and $d^{\kappa\lambda\tau}$, including their specific values and their geometrical interpretation.

\appendix

\section{Triplets of roots summing to zero}\label{app:sum}

Let us first derive the following result obeyed by the defining weights of SU($N$): any subset of $n<N$ defining weights is linearly independent. To prove this result, we can show that the Gram matrix $G_n$ of the scalar products of the vectors in this set is invertible. Owing to Eq.~(\ref{eq:constraints}), this matrix is given by $G_n=(\mathds{1}-1_n/N)/2$, where $a_n$ denotes the matrix with all entries equal to $a$. We now have
\beq
\ln {\rm det}\,(2G_n)={\rm tr}\,\ln \left(\mathds{1}-\frac{1_n}{N}\right)=-\sum_{k=1}^\infty \frac{1}{k}\frac{1}{N^k} {\rm tr}\, (1_n)^k\,.
\eeq
It is easily checked that $(1_n)^k=(n^{k-1})_n$ and thus ${\rm tr}\,(1_n)^k=n^k$. Then
\beq
\ln {\rm det}\,(2G_n) =-\sum_{k=1}^\infty \frac{1}{k}\frac{n^k}{N^k}=\ln\left(1-\frac{n}{N}\right)\,.
\eeq
This shows that $G_n$ is invertible iff $n<N$, as announced. We have also seen in the main text that for $n=N$, the defining weights are constrained by $\sum_{k=1}^N\rho_k=0$ and this is the only possible vanishing combination of the $N$ weights.

Let us now use these two results to show that the only triplets of roots that sum to zero in SU($N$) are of the form $\rho_1-\rho_2$, $\rho_2-\rho_3$ and $\rho_3-\rho_1$. To this purpose, suppose that we have three roots $\alpha=\rho_1-\rho_2$, $\beta=\rho_3-\rho_4$, $\gamma=\rho_5-\rho_6$ that sum to zero but that do not follow the above pattern. We have three different cases depending on the number of distinct weights involved, six, five or four. The first case requires $N\geq 6$  since we need six distinct defining weights. However, the condition $\alpha+\beta+\gamma=0$ writes $\rho_1-\rho_2+\rho_3-\rho_4+\rho_5-\rho_6=0$ which is forbidden both in the case $N=6$, because the only vanishing combination of the six defining weights of SU($6$) is their sum, and in the case $N>6$, since there are no independent sets of six weights for SU($N>6$). In the second case, the roots $\alpha$, $\beta$ and $\gamma$ involve only five distinct roots, which requires only $N\geq 5$. However, up to possible permutations, the only possible configurations are $\alpha=\rho_1-\rho_2$, $\beta=\rho_3-\rho_4$, $\gamma=\rho_1-\rho_6$ or $\gamma=\rho_5-\rho_1$. In the first case, $\alpha+\beta+\gamma=0$ writes $2\rho_1-\rho_2+\rho_3-\rho_4-\rho_6=0$, whereas in the second case, the same condition writes $-\rho_2-\rho_4+\rho_3+\rho_5=0$. None of these are possible since the only vanishing combination of the five defining weights of SU($5$) is their sum and because there are no independent subsets of four defining weights. The third possible case is when the three roots $\alpha$, $\beta$ and $\gamma$ involve only four distinct weights, which requires only $N\geq 4$. Up to possible permutations, the roots should be of the form $\alpha=\rho_1-\rho_2$, $\beta=\rho_1-\rho_3$, $\gamma=\rho_1-\rho_4$ or $\gamma=\rho_4-\rho_1$. In the first case, $\alpha+\beta+\gamma=0$ writes $3\rho_1-\rho_2-\rho_3-\rho_4=0$, whereas in the second case, the same condition writes $\rho_1-\rho_2-\rho_3+\rho_4=0$. None of these are possible since the only vanishing combination of the four defining weights of SU($4$) is their sum. This concludes the proof that, in SU($N$), the only triplets of roots that sum to zero are those associated to a triplet of defining weights.

\end{document}